# Modern Definition of Bioactive Glasses and Glass-Ceramics

Adam Shearer[1], Maziar Montazerian[2], John C. Mauro[1,*]

1- Department of Materials Science and Engineering, The Pennsylvania State University, University Park, Pennsylvania 16802, USA
2- Northeastern Laboratory for Evaluation and Development of Biomaterials, Federal University of Campina Grande, Campina Grande 58429-900, PB, Brazil.

* Corresponding author: John C. Mauro, email: jcm426@psu.edu

**Abstract**

Bioactive glasses (BGs) and glass-ceramics (BGCs) have become a diverse family of materials being applied for treatment of many medical conditions. The traditional understanding of bioactive glasses and glass-ceramics pins them to bone-bonding capability without considering the other fields where they excel, such as soft tissue repair. We attempt to provide an updated definition of BGs and BGCs by comparing their structure, processing, and properties to those of other biomaterials. The proposed modern definition allows for consideration of all applications where the BGs and BGCs are currently used in the clinic and where the future of these promising biomaterials will grow. The new proposed definition of a bioactive glass is "*a non-equilibrium, non-crystalline material that has been designed to induce specific biological activity*". The proposed definition of a bioactive glass-ceramic is "*an inorganic, non-metallic material that contains at least one crystalline phase within a glassy matrix and has been designed to induce specific biological activity*." BGs and BGCs can bond to bone and soft tissues or contribute to their regeneration. They can deliver a specified concentration of inorganic therapeutic ions, heat for magnetic-induced hyperthermia or laser-induced phototherapy, radiation for brachytherapy, and drug delivery to combat pathogens and cancers.

## 1- Introduction: Basic Definitions

*Glasses* have been perhaps one of the most vital materials in the development of human civilization. The glassy state has been recently defined as *"a nonequilibrium, non-crystalline condensed state of matter that exhibits a glass transition. The structure of glasses is similar to that of their parent supercooled liquids (SCL), and they spontaneously relax toward the SCL state. Their ultimate fate, in the limit of infinite time, is to crystallize"* [1]. Several factors differentiate glasses from other materials: glasses are thermodynamically unstable and only exist under the glass transition temperature causing spontaneous, continuous relaxation [2]. These materials have found their way into several applications ranging from semiconductors to biomaterials. The glass structure and its relationship with properties is a critical distinction from crystalline materials. For example, the chemical dissolution of BGs and BGCs could be tailored through the network connectivity, the choice of glass network former, and more advanced phenomena such as exploitation of the mixed alkali effect. Glass degradation releases a cascade of inorganic ions, which have since been found to host a battery of therapeutic effects on the human body [3].

An updated definition of *glass-ceramics* reads, "*glass-ceramics are inorganic, non-metallic materials prepared by controlled crystallization of glasses via different processing methods. They contain at least one type of functional crystalline phase and a residual glass. The volume fraction crystallized may vary from ppm to almost 100%*" [4]. The particular chemical composition of glassy and crytalline phases, as well as their nanostructures or microstructures, has resulted in a wide range of remarkable properties and applications in the fields of domestic, defense, space, electronics, health, architecture, energy, chemical, and waste management [5].

A *bioactive material* is "*a material which has been designed to induce specific biological activity*." The definition that achieved consensus in Chester, UK 1986 [6]. The experts in a consensus conference on the definition of biomaterials held in Chengdu, China, in 2018 have reaffirmed it. Also, *bioactivity* reads, "*phenomenon by which a biomaterial elicits or modulates biological activity*" [7].

In 1969, Larry L. Hench sparked a revolution in biomaterial research by discovering *bioactive glasses*, the first synthetic device showing bone bonding abilities. Through the dissolution of critical ions, oversaturation fosters the precipitation of hydroxyapatite at the glass/host-tissue interface, creating a bond. BGs have been proven successful devices in the clinic and are now an essential material in the discussion on tissue engineering. Early applications of bioactive glasses relied on this unique ability to stimulate bone bonding with the implant. Bioglass®-EPI (extracochlear percutaneous implant) and endosseous ridge maintenance implant (ERMI®) were monolithic glass implants used in the middle ear and endosseous ridge, respectively. These first-generation materials thus influenced the old definition of bioactive glass [8].

Several compositions of bioactive glasses have been used, relying primarily on silicate-based compositions but also borate and phosphate glasses [9]. New glass network formers have been explored as the applications expanded beyond bone regeneration. Borate and phosphate-based compositions have shown faster dissolution and more suitable bioactivities for soft tissue applications. Compositional innovation has broadened the application of BGs, allowing for advanced biological properties such as nerve regeneration and cancer treatment. Table 1 lists the most common bioactive glass compositions seen in the literature.

**Table 1.** Most researched bioactive glass compositions in wt.%.

| | |
|---|---|
| **45S5** | $45SiO_2–24.5Na_2O–24.5CaO–6P_2O_5$ |
| **13-93** | $53SiO_2–6Na_2O–12K_2O–5MgO–20CaO–4P_2O_5$ |
| **13-93B3** | $53B_2O_3–6Na_2O–12K_2O–5MgO–20CaO–4P_2O_5$ |
| **S53P4** | $53SiO_2–20CaO–23Na_2O–4P_2O_5$ |
| **70S30C** | $70SiO_2–30CaO$ |
| **58S** | $58SiO_2–33CaO–9P_2O_5$ |
| **1-98** | $52.7SiO_2–1B_2O_3–6Na_2O–11K_2O–5MgO–22CaO–2P_2O_5$ |
| **P50C35N15** | $71P_2O_5–19.7CaO–9.3Na_2O$ |

**2- Glass Structure**

Oxide glasses, such as the bioactive glasses listed in Table 1, have network structures primarily composed of $SiO_4$ tetrahedral building blocks or various $BO_x$ and $PO_x$ structural units [10]. These structural units are connected through mixed ionic-covalent bonding being disrupted by forming non-bridging oxygens (NBO). These oxide glasses differ from other glasses, including chalcogenides, metallic glasses, and organic polymeric glasses. Chalcogenides are non-oxide glasses containing one or more chalcogens (sulfur, selenium, or tellurium) and are built from covalently bonded structural units. Despite their lack of medicinal relevance, they have widespread applications in other fields. Metallic glasses consist of icosahedral units with metallic bonding. Although they have not been studied as extensively, bulk metallic glasses show some bioactive properties and should be considered under the definition of bioactive glasses as they may have a greater significance in future research [11]. Organic polymeric glasses are made through cross-linking of molecular chains bonded through van der Waals forces and covalent intra-molecular bonding. No studies have shown the bioactive potential of these organic glasses, but they should not be removed from the definition of BGs. They may each show promising novel properties once extensive research on biocompatibility, biodegradability and biological stimulation has been considered. Glasses exhibit an unlimited range of structures, states, and behaviors. It is estimated that more than $10^5$ different types of inorganic glasses have already been studied, but there is still room for billions more to be developed [12].

Glasses do not exhibit long-range order, as seen in most other crystalline structures such as the ceramic precursor materials used to form glasses. Instead, these materials have a disordered long-range structure with some short to medium ranged ($3^{rd} - 4^{th}$ nearest neighbors) [10]. The most widely used bioactive glasses consist of $SiO_2$, $B_2O_3$, and $P_2O_5$ as glass network formers with alkali and alkaline earth elements as network modifiers. There is an extensive catalogue of experimental and computational data, theoretical calculations, and systematic reviews on glass structure. It is a continual interest in solid-state chemistry, condensed matter physics, geosciences, and medicine [13].

Silicate glasses are crucial for many technologies, e.g., optical fibers, electronic screens, architectural materials, and bioactive glasses, but they also exist in nature. Crystalline silica comprises three-dimensionally linked, corner-sharing networks of $SiO_4$ tetrahedra (Figure 1a). Oxygen anions link the $Si^{4+}$ cations through –Si–O–Si– networks and are considered bridging oxygens (BO). The addition of alkali or alkaline earth oxides to a silica melt increases the number of oxygen anions making the tetrahedral coordination thermodynamically unfavorable. As a result, some of the $1^+$ or $2^+$ cations are forced to participate in bonding with oxygen in the $SiO_4$ structural unit. This oxygen is then determined "non-bridging" as it does not contribute to the polymerization of the network and disrupts the structure. A two-dimensional visualization of a stereotypical Na$_2$O–CaO–SiO$_2$ glass composition is shown in Figure 1b.

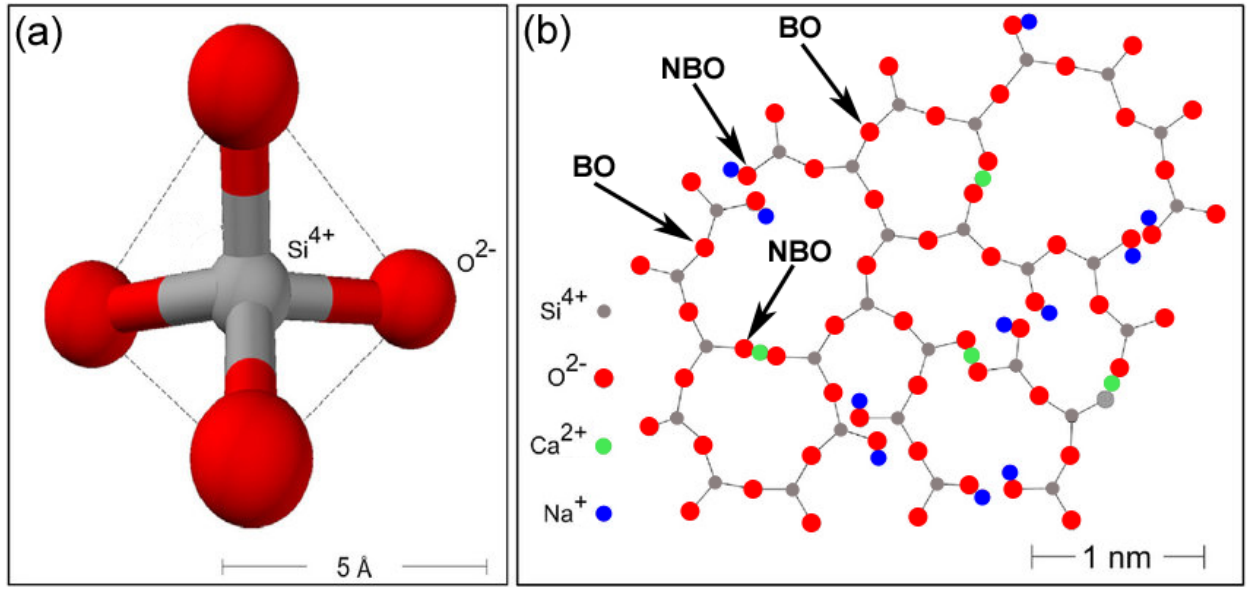

**Figure 1**. Model of (a) SiO$_4$ tetrahedra and (b) two-dimensional visualization of a typical soda lime silicate glass network structure comprising SiO$_4$ tetrahedra, bridging (BO) and non-bridging oxeyes (NBO), connected with Na$^+$ and Ca$^{2+}$ cations. Adapted from Maraghechi *et al.* with permission from Elsevier [14].

Unlike silicate glasses, borate glasses have two distinct structural units that can form due to coordination with oxygen, $^{III}$B and $^{IV}$B. Pure B$_2$O$_3$ glasses show only the lowest coordinated state while introducing alkali or alkaline earth shifts the average coordination number. As an alkali oxide is added, the coordination of the boron increases, thus showing a higher concentration of $^{IV}$B structural units. The [BO$_{4/2}$]$^-$ structural unit has an excess negative charge which is in turn balanced by an alkali ion. The boron anomaly refers to the behavior of borate glasses with increasing alkali content. Adding alkali to a borate glass causes a shift from BO$_3$ to BO$_4$ without causing the creation of NBOs [15]. This shift in coordination without the formation of NBOs increases the connectivity of the glass network, which in turn increases the viscosity and decreases thermal expansion coefficient.

The glass-forming component of phosphate glasses is $P_2O_5$. The dominant structural unit of phosphate glasses is the orthophosphate ($PO_4^{3-}$) tetrahedron, which consists of a $P^{5+}$ cation surrounded by four $O^{2-}$ anions. One oxygen in this structure is double-bonded terminal oxygen, while the other contributes to the polymerization of the network as BOs. Bridging these oxygens links the individual phosphate tetrahedra through covalent –P–O–P– bonds. Phosphate glasses can have various structural units ranging from $Q^3$ to $Q^0$ showing decreasing amounts of bridging oxygens due to network modifier addition. The level of polymerization can also determine the medium-order structure by forming chains, rings, or branching structures. Increasing the network modifier concentration disrupts the phosphate glass structure by creating ionic cross-linkages between NBOs [16].

Cation field strength is used to generally describes the effects of different cations on the properties and structure of glasses, e.g., the dissolution kinetics of BGs [17]. The general formula is the valence of the cation ion divided by the sum of the radii of the cation and anion. For example, cation field strength would be used to compare cations with varying radii, such as $Mg^{2+} > Ca^{2+} > Na^+ > K^+$. When multiple cations are combined, the higher field strength ion dominates, creating small coordination shells with short bond distances. Combining several glass network modifiers will increase the configurational entropy, affect properties such as viscosity and thermal expansion, and change other thermodynamic variables. Randomized distribution of the modifiers in the glass network largely contributes to properties [18]. A 3D model of a bioactive yttrium aluminosilicate glass ($Y_2O_3$–$Al_2O_3$–$SiO_2$) illustrates the possibilities for disorder in a glass structure (Figure 2) [19]. This compositional family is used in *in situ* radiotherapies to irradiate certain cancer tumors with radioactive $^{90}Y$. This radioactive cation changes the coordination and properties of glasses and melts in intricate ways, making it exceptionally relevant to understand structure-property relationships in glasses.

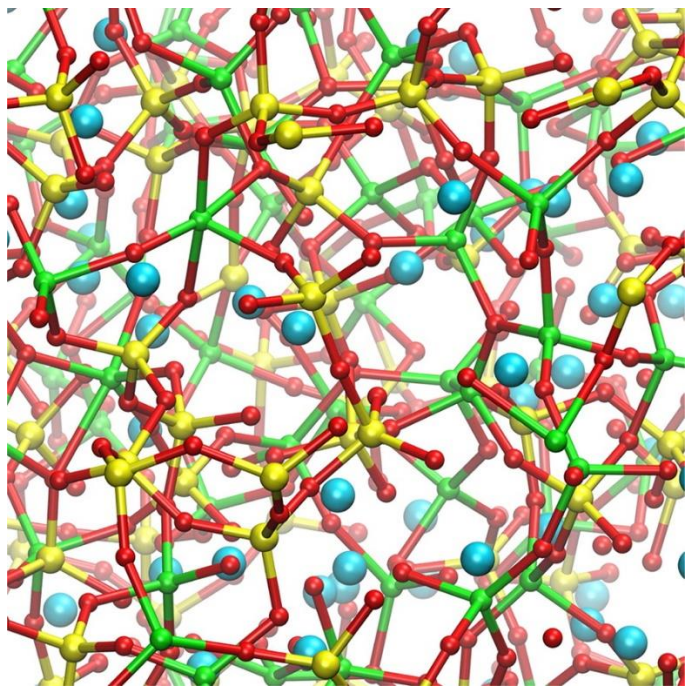

**Figure 2**. Ball and stick model of an yttrium aluminosilicate glass structure. The yellow balls represent silicon atoms, the green represents aluminum atoms, the blue represents yttirum, and lastly red represents oxygens. Reprinted from Christie and Tilocca with permission from ACS [19].

Glass-ceramics which are obtained by controlled crystallization of BGs have also received great interest in the field of biomaterials with several commercial products introduced to the market (e.g., CeraBone®, BioSilicate®, BioMin®, etc.). Based on the definition of glass-ceramics, some BGs that undergo unwanted crystallization after melt-quenching, stabilization of gels, or sintering scaffolds might be considered bioactive glass-ceramics. However, controlled crystallization is always preferable to make the process scalable. Controlled crystallization of a biocompatible secondary phase strengthens the matrix without sacrificing much bioactivity. The most popular crystalline phases showing biocompatibility include apatite, wollastonite, mica, and combeite. The commercial state and rising compositions in BGCs have been thoroughly reviewed by Montazerian and Zanotto [20]. For example, Peitl *et al.* showed that the controlled crystallization of the glass would increase the 4-point bending strength by nearly a factor of 3 compared to uncrystallized parent compositions [21]. As a result, the mechanical properties of BioSilicate® glass-ceramic show a maximum flexural strength at 40 Vol% of crystallized sodium calcium silicate bioactive phase ($Na_2CaSi_2O_6$). Figure 3 further visualizes BGC mechanical properties, showing 4-point bend test results with relevant optical microscopy images. This BGC exhibits the best mechanical performance without sacrificing biocompatibility among the current commercial BGC products.

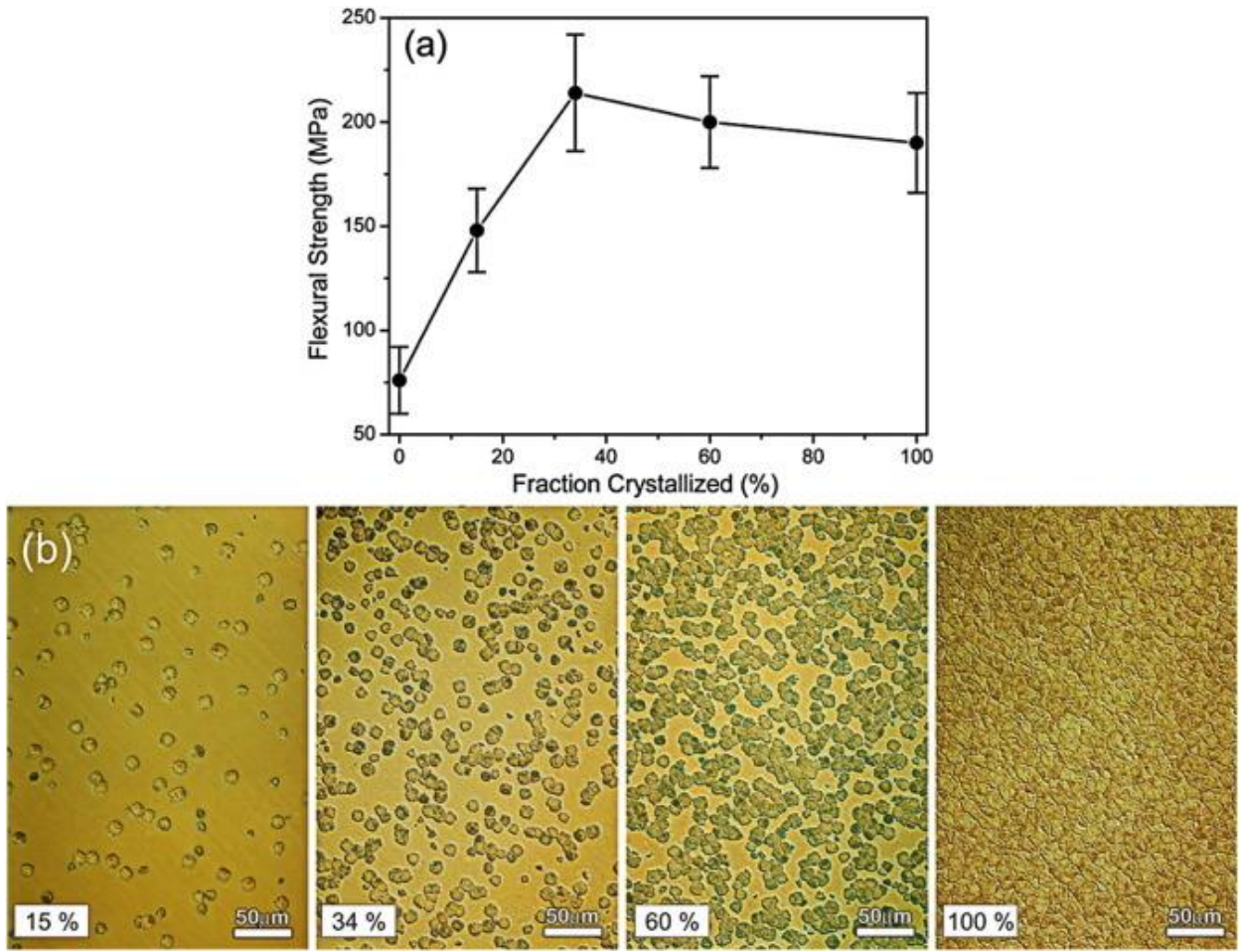

**Figure 3**. (a) 4-point bend test strength results versus crystalline volume fraction for the BioSilicate® bioactive glass-ceramic composition. (b) optical microscopy images showing the microstructure for partially to fully crystallized glass-ceramics. Reproduced from Peitl *et al.* with permission from Elsevier [21].

**3- Processing**

Different processing methods exist to formulate BGs or BGCs depending on the desired clinical application. BGs can be synthesized through the traditional melt-quenching technique or sol-gel method. The sol-gel process allows for greater physical and chemical control of the final product allowing for a highly tailorable product. The processes for both melt-quenched and sol-gel synthesized glasses are summarized and illustrated in Figure 4. The sol-gel synthesis route demonstrates better control over microstructure, homogeneity, and particle size [22]. Moreover, the microporosity and high specific surface area of gel-derived glasses stimulate rapid dissolution and accelerated hydroxyapatite formation at the surface of the glass as a result of the high surface area and micro/nano porosity [23].

The melt-quenching approach requires high temperatures (1300-1600 °C) to melt precursor reagents and homogenize them. Proper melting procedures (crucible type, heating rate, dwell time, etc.) are required to produce a homogenous, defect-free product which is required for a high-quality medical device. Melt-derived glasses (or melt-glasses) can then be cast, fritted, or drawn into several different morphologies [24]. Monolithic pieces, fritted glass, and fibers have all been studied and implemented into the clinic as BG devices. The sol-gel process has been defined as a wet chemistry-based processing technique at low temperatures for the production of ceramic materials. The 3D glass network forms at room temperature via the polymerization reaction of a precursor solution with chosen reagents based on the desired composition and intended application. Sol-gel production of BGs has three main steps: the preparation of the precursor solution (sol), the gelation of the sol, and the removal of solvents and salts through a thermal treatment which also results in structural stabilization [25]. Gel-derived glasses (or gel-glasses) can easily be doped with the introduction of therapeutic inorganic ions based on the desired biological effect (angiogenesis, antibacterial or anti-inflammatory effects, or drug delivery applications) [26]. Doping a composition is much easier via the sol-gel process than the melting process.

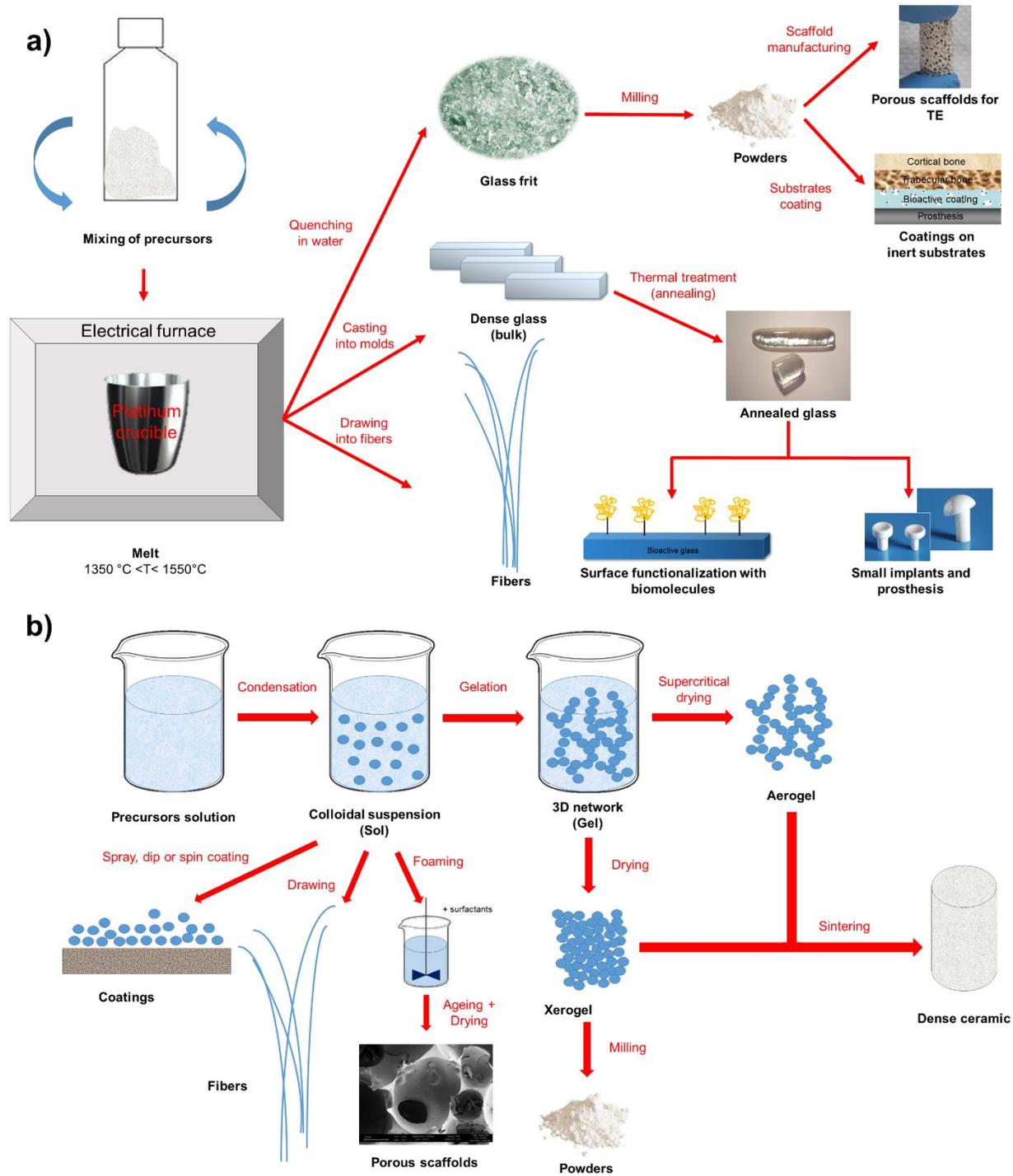

**Figure 4**. Schematic representation of (a) melt-quenching and (b) sol-gel route for bioactive glass synthesis and final products [27].

The final product of the sol-gel process can be in many different morphologies, including monoliths, porous scaffolds, fibers, coatings, or granules. Some are characterized by the mesoporous texture inherent to the sol-gel and self-assembly processes [23]. These sol-gel materials can also be used to produce nanoparticles to be used as drug delivery carriers for drug loading and release in the treatment of bone pathologies, cancer treatment, or medical imaging applications [27,28].

Glass-ceramics are produced through carefully controlled thermal treatments that provide sufficient energy to overcome the thermodynamic barriers for crystallization. Thermal characterization techniques such as differential scanning calorimetry (DSC) must determine the proper temperatures for these heat treatments. A recent study by Fiume *et al.* compares the physical properties of BGs and BGCs derived from both melt-quenching and sol-gel synthesis [29]. The same composition was used across all processing methods to target bone tissue regeneration. Gel-glasses were determined to have a specific surface area 2-4 times greater than melt-glasses and slightly less for the gel-BGCs. Controlled crystallization of gel-derived glass-ceramics is also possible through controlled heat treatment on already calcined gels in the form of both monoliths and powders. However, this approach requires additional steps to produce the gel-glasses first.

More cutting-edge methods for making and processing BGs, like 3D printing, sputter coating, and chemical vapor deposition, are being developed. Several studies have embedded BG powder into polymer precursor materials used in 3D printing [30–32]. Embedding BG powder into the 3D-printed polymers, injectable hydrogels, and cements increases the bioactivity of the devices allowing for greater osteointegration and handleability. Several 3D printing techniques, such as selective laser sintering, direct ink writing, and stereolithography, have all been studied and reviewed [33]. Thin film BGs have been deposited onto metals used as implant materials in an attempt to increase the osteointegration and bond strength of the implant-bone interface. BG coatings have been successfully applied to various metals and polymers, including medical-grade titanium, magnesium alloys, a variety of stainless steels, PMMA, etc. [34].

## 4- Properties

### 4-1- The Conventional Understanding

Bioactive glasses were originally defined by their ability to precipitate hydroxyapatite and bond with the surrounding bone tissue. Hench originally defined the term bioactive as the following, *"a bioactive material is one that elicits a specific biological response at the interface of the material which results in the formation of a bond between the tissues and the material"* [35]. The chapter on BGs continues to narrowly characterize them as melt-derived with bone-bonding characteristics. The original phase diagram created by Hench in 1991 shows the greatest bioactivity where the most amount of bone bonding was found (Figure 5). The bioactivity index ($I_B$) is determined in terms of the time (in days) it takes for a BG to bond with half of the implant surface *in vivo*. BGs and BGCs were quickly found to stimulate biological responses beyond the formation of an apatite layer. As such, they have been dubbed "third-generation biomaterials" with the goal of functional tissue regeneration. Biomaterials of the third generation are molecularly targeted stimulators of specific cellular responses; they are also bio-interactive, bio-integrative, and resorbable. A critical factor of these materials is the ability to activate and upregulate several critical growth factors, transcription factors, cell cycle regulators, apoptosis regulators, cytokines, and extracellular matrix compounds [36].

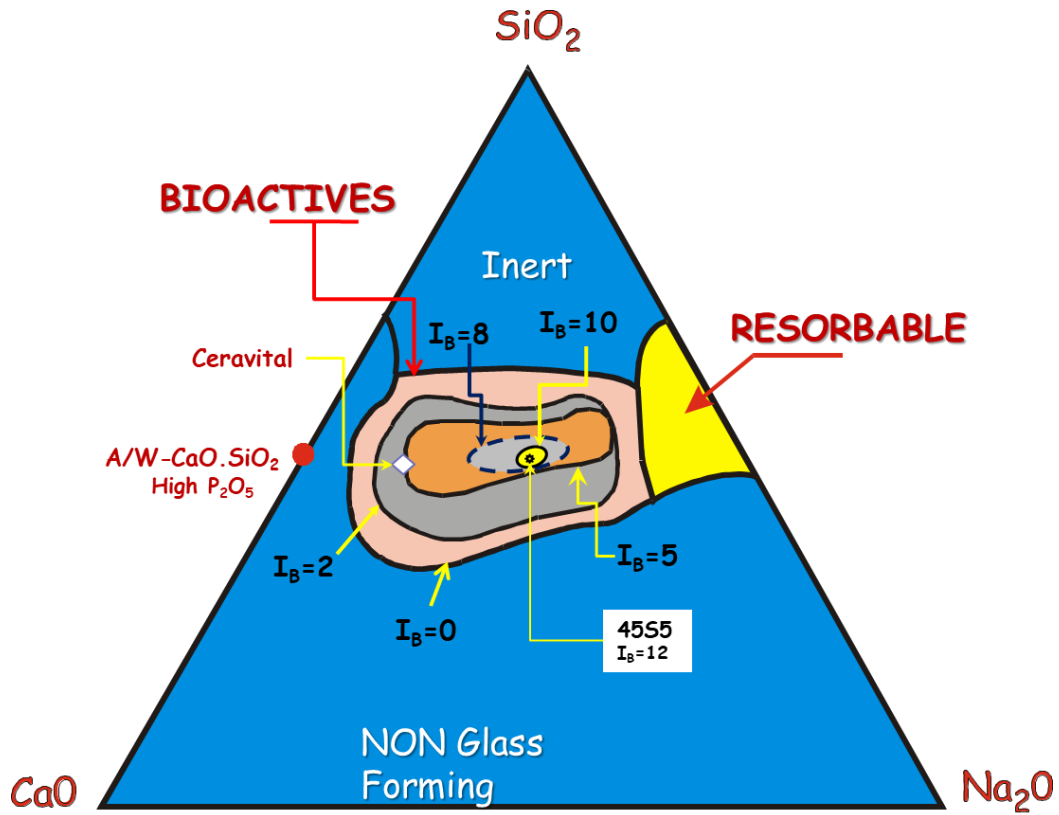

**Figure 5**. Hard tissue and soft tissue bonding ability as a function of composition in the $SiO_2$–$Na_2O$–$CaO$–$P_2O_5$ system while also showing the bioactivity index of each phase, reproduced with permission from Refs. [37,38].

### 4-2- Motivation for New Definition

BGs have expanded applications outside hard tissue regeneration, showing multiple morphologies, applications, and compositions where hydroxyapatite conversion is not a determining factor. BGs have become dominant in the conversations around soft tissue regeneration and wound healing due to their ability to stimulate genetic upregulation. Therapeutic ions release and promote reepithelization and formation of new tissue, making BGs a prime candidate for wound healing devices. Other applications show preliminary data promising BGs-based devices for use in nerve/muscle regeneration, drug delivery and cancer therapies [39–43].

BGs and BGCs now have several desirable attributes to stimulate different biological processes. Osteogenic elements can assist in the signaling of proteins to suppress osteoclastic activity or promote osteoblasts to regrow bone. Angiogenic elements aid in reepithelization, forming extracellular matrices, and reforming damaged blood vessels. Some elements have been shown to provide therapeutic effects against cancer cells or have been used to aid in medical imaging through photoluminescence effects. Lastly, several transition metals have been shown to act as antipathogenic agents, helping prevent the growth of gram-positive and gram-negative bacteria, viruses, and fungi. Figure 6 provides a comprehensive visualization of the periodic table and how each element assists in the applications of bioactive glass. This table shows several elements studied for their antimicrobial properties, focusing primarily on transition metals. These ions kill harmful microbes in the physiological environment through several mechanisms, including producing reactive oxygen species, cell membrane dysfunction, protein dysfunction, or genotoxicity [44]. Silver and copper appear commonly among the most studied ions as they kill both gram-positive and gram-negative bacteria. The anticancer abilities of BGs and BGs have been reviewed thoroughly for their ability to deliver therapeutic drugs or irradiate cancerous cells [45–47]. Cancerous cells can also be treated through photothermal effects or magnetic hyperthermia [39]. The mechanical properties can be improved by using components that are more commonly found in glass-ceramics or strengthening composites, such as cements or dental resin composites [48]. These applications require stringent mechanical properties similar to the surrounding material (most commonly bone or teeth) to perform properly.

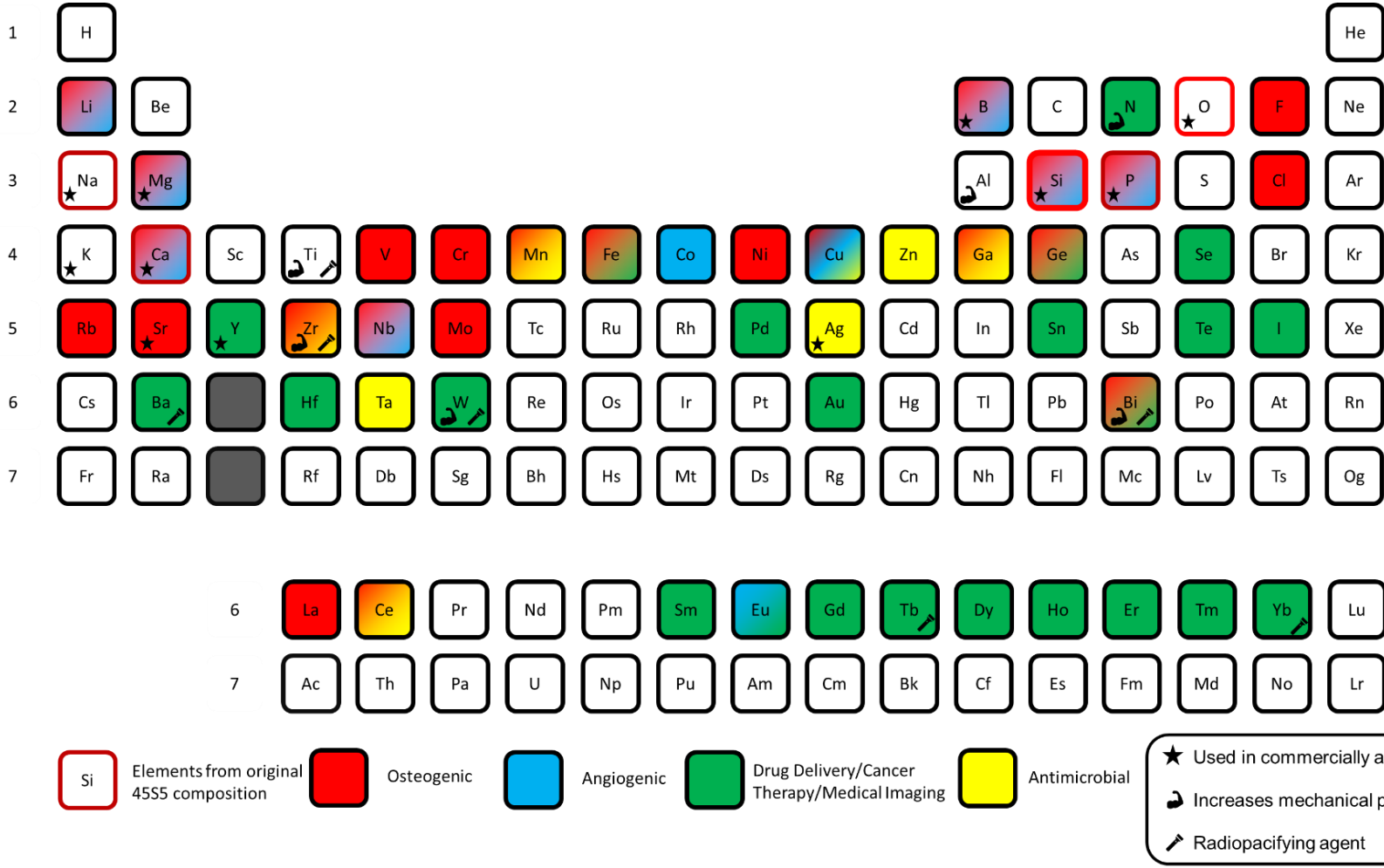

**Figure 6**. Periodic table of the elements used in bioactive glasses labeled based on their biological and structural properties. Elements with a red border signify those which are used in the original 45S5 Bioglass® composition. Each of the colored elements represent a different biological application where some show multiple beneficial properties. Labels in the corners of the elements signify which are used in commercial products, increases the glasses mechanical strength, or act as a radiopacifier.

## 5- Proposed Modern Definition

The new proposed definition of a bioactive glass is "*a non-equilibrium, non-crystalline material that has been designed to induce specific biological activity*." The proposed definition of a bioactive glass-ceramic is "*an inorganic, non-metallic material that contains at least one crystalline phase within a glassy matrix and has been designed to induce specific biological activity*." BGs and BGCs have the unique ability to deliver a specified concentration of inorganic therapeutic ions. They can stimulate or aid in several advanced medical processes, including providing heat for magnetic-induced hyperthermia or laser-induced phototherapy, radiation for brachytherapy, and delivery of critical proteins and drugs. The several applications and clinical uses of bioactive glasses are illustrated in Figure 7. It is evident that dozens of commercial potential clinical products are possible in the fields of bone regeneration, soft tissue repair, dentistry, medical imaging, cancer therapy, and drug delivery.

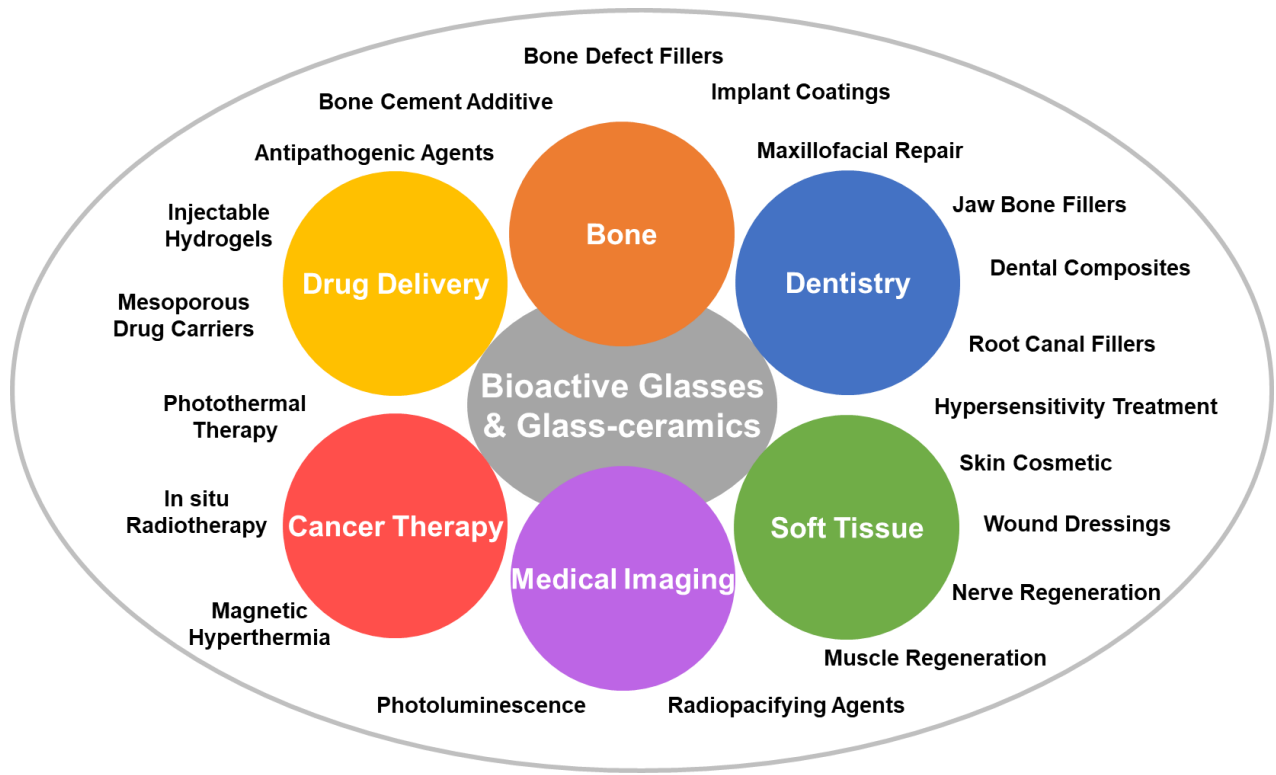

**Figure 7**. Illustration highlighting the several applications of bioactive glasses and glass-ceramics.

**Conclusions**

The future of bioactive glasses is diverse, showing broadened compositional spaces with new therapeutic inorganic ions, advanced processing methods, and a continuously evolving and expanding range of medical applications. Innovation has produced novel morphologies, sub-micron particle sizes, and a catalogue of excipients for BG delivery. Although commercial BG and BGC products have favored hard tissue repair, a new wave of products is expanding into soft tissue regeneration, drug delivery and cancer treatment. The new proposed definition allows for each of these novel biomaterials to be included and better represents the field of BGs and BGCs to all communities.